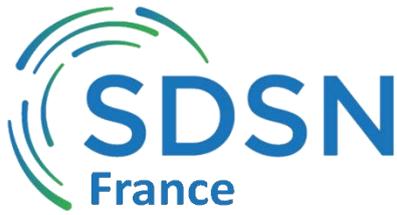
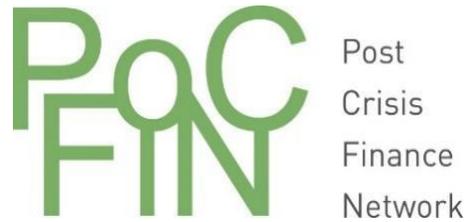
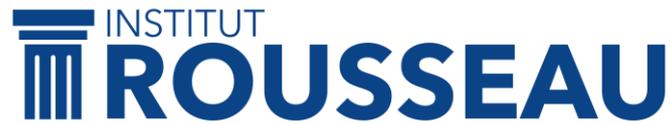

# Policy Brief Agenda 2030
## 2021/09

## Repenser le financement des entreprises vertueuses et les politiques prudentielles en intégrant la solvabilité socio-environnementale

par

## Camille Souffron[1] & Laura Chémali[2]

[1] Etudiant à Paris-Saclay, Paris 8, Paris 10 et Montpellier 3. Email : camille.souffron@cri-paris.org
[2] Analyste au sein d'une agence de notation socio-environnementale. Email: laura.chemali@gmail.com

**Introduction**

Face aux contraintes environnementales de plus en plus pressantes[3] et à **l'insuffisance de l'effet du *pricing* du carbone**, une transition environnementale va nécessiter des **investissements massifs "vertueux" vers les secteurs décarbonés et une réduction des investissements dans les secteurs polluants, mais aussi dans les secteurs sociaux, solidaires et circulaires.** Or, les investissements verts sont encore largement insuffisants. En effet, leurs coûts initiaux étant généralement très importants, il est rare que les entreprises aient les fonds propres nécessaires, et ces dernières se tournent donc vers des sources de financement externe (crédit bancaire, marchés de capitaux, marchés boursiers) et se heurtent à une **indisponibilité des ressources financières.**

Le crédit bancaire étant souvent rationné, les entreprises sont donc incitées à chercher des financements sur les marchés de capitaux et boursiers, difficilement trouvables du fait de faibles rendements et liquidité et de hauts risques, auxquelles s'ajoutent les appels de marge des chambres de compensation[4]. En prenant des appels de marge, elles garantissent la transaction si l'une des parties fait défaut, afin d'éviter une faillite à la chaîne. Or, la hauteur des appels de marge et les commissions des courtiers peuvent pousser des entreprises à quitter les marchés organisés eux-mêmes pour se financer **sur les marchés de gré à gré risqués et dark pools**, augmentant le risque systémique. Trois problématiques en découlent :

- La **dégradation de l'environnement** et son impact sur la production ;

- Le **risque systémique financier** du fait de la dérégulation financière et de la croissance des marchés de gré à gré ;

- Le **risque macro-économique déflationniste :** une explosion de la dette privée et une situation économique trop faible pour la soutenir mènent à une spirale déflationniste et à une dépression forcée (scénario "à la japonaise"). Le niveau de dette privée provoque une faillite des ménages et des entreprises en chaîne et donc potentiellement des banques, une explosion du chômage, et bloque tout investissement.

Nous en concluons qu'il semble impossible de mener une politique environnementale qui ne soit pas macro-prudentielle, car il ne serait alors pas possible de réunir les incitations et investissements nécessaires à la transition. Et inversement, une politique macro-prudentielle doit nécessairement être environnementale[5], autrement la dégradation de l'environnement mènerait à la dégradation de la production, du capital, de l'emploi, du ratio d'endettement et de la qualité de vie sur terre, et donc à une dépression forcée. **Les politiques environnementales et macro-prudentielles sont donc liées[6], et il est vital que cette interdépendance soit comprise par le régulateur** avec la réalisation de *stress-tests* environnementaux**.**

Nous proposons donc une **politique macro-prudentielle *écosystémique*** intégrant la régulation socio-environnementale, dont l'objectif serait d'**augmenter les capacités de financements, les incitations et l'accès à l'investissement vert tout en réduisant les risques déflationnistes et environnementaux.** Elle se décline à la fois sur le financement bancaire, le refinancement des banques commerciales et le financement de marché désintermédié, pour cibler tout type d'entreprise et toute source d'instabilité financière, en s'appuyant sur une redéfinition des principes comptables intégrant la solvabilité socio-environnementale.

---

[3] World Climate Research Program. (2019), *Rapport de la 22ème Session de Travail - CMIP6.*
[4] Une chambre de compensation est un organisme financier qui centralise les transactions sur les marchés dits organisés.
[5] Le but d'une régulation macro-prudentielle est la limitation du risque financier systémique.
[6] Bovari, E., Giraud, G., & Mc Isaac, F. (2018). Coping With Collapse : A Stock-Flow Consistent Monetary Macrodynamics of Global Warming. *Ecological Economics*, *147*, 383-398.

**Politiques de régulation macro-prudentielles écosystémiques**

L'objectif de ces politiques est d'élaborer un système de réallocation des capacités de financement des entreprises non-vertueuses vers les plus vertueuses avec des garanties publiques, visant à réduire le ratio d'endettement tout en augmentant les investissements verts, avec des politiques monétaires et des formes de partenariat public-privé pour ce faire. **Faciliter le financement des entreprises vertueuses** verdirait le capital total mais augmenterait son volume, neutralisant en partie l'impact environnemental positif. Il est donc nécessaire d'également **limiter l'expansion du financement des entreprises "brunes"**. Cela réduirait les opérations risquées et favoriserait des investissements à effet de levier plus faibles et plus connectés à l'économie réelle, diminuant le risque financier systémique.

Des régulations prudentielles (Bâle III, Solvabilité II) ont déjà été introduites après la crise de 2007, obligeant les banques à maintenir un certain ratio de liquidité et de solvabilité[7] afin d'éviter une nouvelle crise causée par une expansion incontrôlée du crédit et une crise de liquidité. Toutefois, ces régulations absolument nécessaires viennent directement pénaliser l'investissement vert car plus risqué et peu rentable.[8] L'enjeu est de **ne pas pénaliser les investissements vertueux tout en maintenant un même niveau prudentiel écosystémique** justement par la diminution du risque environnemental et le transfert des surcapacités de financement, garanties et collatéraux des investissements polluants vers les investissements vertueux. Ainsi, plusieurs solutions existent pour diminuer la contrainte bancaire de crédit :

● *Une politique des taux*

Bien que les trois principaux taux directeurs, communs à toute banque, soient déjà faibles, il serait possible de **les différencier selon les établissements et les prêts, en** :

- Individualisant le taux de refinancement auprès de la banque centrale (refi) pour chaque établissement et demande de refinancement, avec un taux plus faible pour les banques avec un bilan/une demande "verts", et un taux réellement pénalisant pour les bilans/demandes "bruns".

- Réduisant et plafonnant le taux de prêt marginal (le taux de prêt de liquidité aux banques commerciales qui dépassent le volume de création monétaire permise par le refi, prêt toujours accepté) pour les banques à bilan vert, et inversement pour les bilans bruns.

- Augmentant le taux de rémunération des dépôts à la banque centrale seulement pour les banques qui ont un niveau particulièrement élevé de créances vertes. Cela inciterait les banques à verdir leur bilan pour profiter de cette rémunération aujourd'hui négative donc à perte, et une fois verdis les dépôts réalisés à la banque centrale limiteraient d'autres effets de levier et auraient un rôle prudentiel.

Ainsi, individualiser les taux pour chaque demande de refinancement serait particulièrement incitateur pour les banques et renforcerait la transparence d'information pour le régulateur car obligerait les banques à motiver leurs demandes selon le type de prêt et à flécher leur allocation de crédit. **Pour**

---

[7] Maintenir un certain ratio de liquidité signifiant disposer d'assez d'actifs liquides comme obligations d'Etat, fonds propres, réserves à la banque centrale, pour garantir un retrait massif des épargnants ou des faillites des débiteurs.
[8] Spencer, T., Stevenson, J. (2013). "EU Low-Carbon Investment and New Financial Sector Regulation: What Impacts and What Policy Response?". *Working Papers n°04/13*, IDDRI, Paris, France, 18 p.

**différencier ces taux, les banques centrales devraient accepter les collatéraux (les garanties) verts** comme sûretés, ce qui nécessite des normes et évaluations rigoureuses des entreprises débitrices.

- *Une politique des ratios de réserves et fonds propres*

    Il est aussi possible de **jouer directement sur les ratios de réserves et de fonds propres**, donc sur les volumes contraignant le crédit et la création monétaire. Cependant, la majorité des banques centrales des pays développés (BCE, Fed…) mène avant tout des politiques de taux afin de contrôler et harmoniser le taux de prêt interbancaire[9], et acceptent ce faisant toute demande de refinancement. En les refusant en présence de tensions et d'illiquidité, elles provoqueraient des variations du taux interbancaire, rendant les taux hors de leur contrôle. Néanmoins **il est possible d'imaginer une plus grande volatilité des taux, et un contrôle directement par les réserves et volumes**, comme la Banque Populaire de Chine (BPC), offrant alors une myriade d'outils déjà utilisés par le passé en Europe[10].

- Allègement des ratios de liquidité et solvabilité[11] pour les banques à bilan vert et pour les types de prêts accordés, augmentation pour les autres, selon la taille des établissements[12]. Création d'un facteur de pénalité pour les banques finançant des activités polluantes fossiles, en exigeant une hausse des fonds propres proportionnelle au crédit octroyé. Cela permettrait de compenser l'allègement des fonds propres pour les crédits aux activités vertes. Pondération différente du ratio de solvabilité par le risque selon le type de créance. **Une créance verte connaîtrait une baisse de son risque total par son faible risque socio-environnemental, l'Etat se portant alors garant de sa solvabilité financière.**

- Allègement des réserves obligatoires à déposer à la banque centrale pour les banques à bilan vert et pour les types de prêts accordés, augmentation pour les autres. Les obligations vertes seraient acceptées et auraient plus de valeur et de poids que les traditionnelles obligations d'Etat.

- Pondération du volume du coussin prudentiel contra-cyclique dans les fonds propres selon le risque socio-environnemental, aujourd'hui très faible (0%)[13] et fixé par les autorités nationales et non la BCE.

- Des **politiques de *Green Quantitative Easing*** sont également envisageables : la création monétaire et l'injection de liquidités directement par le rachat par la banque centrale non plus seulement des obligations d'Etat mais des obligations vertes. Une politique de *Green QE* serait pleine de sens en terme prudentiel, d'autant plus face à la COVID-19 : le *QE* actuel, en inondant les marchés de liquidités, fait augmenter drastiquement depuis 2009 le prix des actifs boursiers alors même que la situation économique est morose depuis 10 ans. Cette décorrélation entre la valeur des actifs et leur performance économique réelle très faible (cf. *Price Earning Ratio*) indique que les marchés sont sous perfusion des banques centrales, menant à d'importants risques systémiques en cas d'arrêt de la perfusion. Ainsi, un *Green QE* créerait toujours des liquidités et inciterait à l'investissement mais dans les secteurs nécessaires, tout en **limitant les effets de levier** actuels purement spéculatifs ainsi que l'investissement dans des produits financiers dangereux.

---

[9] Taux auquel les banques commerciales se prêtent entre elles.
[10] Lim, C., Columba, F., Costa, A., Kongsamut, P., Otani, A., Saiyid, M., Wezel, T., Wu, X. (2011). "Macroprudential policy: what instruments and how to use them? Lessons from country experiences". *IMF Working Paper.* International Monetary Fund, Washington D.C.
[11] Fonds propres minimums équivalents à 8% des dépôts pondérés par le risque, pour faire face à une perte de confiance et un retrait d'une partie des épargnants.
[12] Rozenberg, J., Hallegatte, S., Perrissin-Fabert, B., Hourcade, J.-C. (2013). "Funding low-carbon investments in the absence of a carbon tax". *Clim. Pol. 13,* 134–141.
[13] Fixé par le HCSF depuis le 2 avril 2020.

- Enfin, il serait aussi pertinent de réduire les obligations de fonds propres de Solvabilité II concernant les investissements en infrastructures souvent nécessaires aux futurs investissements verts privés, obligations bien trop élevées et injustifiées pour ce type d'investissement structurel de long terme.

Dans le cadre de l'Union Européenne, les **décideurs nationaux ont tout intérêt à construire un rapport de force face à la BCE** dans l'objectif d'un tel contrôle monétaire vert, et d'une utilisation des politiques de ratio en plus de celles de taux.

**Partenariat Public/Privé (PPP)**

Le financement est aussi possible non pas seulement par création monétaire mais par réallocation d'actifs. Sur les marchés financiers, des acteurs non-bancaires agissent aussi, les **"investisseurs institutionnels"[14]**, responsables du *shadow banking*. Esquivant les régulations bancaires et prudentielles post-2008, ils sont sources de risques systémiques énormes. L'Etat a toute sa place pour réglementer et créer des incitations dans ce secteur et dans les transactions de marché. Des appels de marge préférentiels dans les transactions en chambres de compensation pour les opérations vertes, et une hausse des marges pour les transactions de pollueurs pour compenser et garder au même niveau leur réserve de résolution et de garantie, pourraient inciter à sortir des marchés de gré à gré et faciliter les opérations vertes.

L'impact macro-prudentiel d'une telle politique serait fort : par la baisse des appels de marge, elle crée une incitation conséquente pour une part des entreprises à se retirer des dark pools et du marché de gré à gré non-couvert, pour se financer dans les marchés couverts et régulés. Et **par la baisse des taux d'intérêt et la hausse des capacités de financement bancaires, elle en incite d'autres à se retirer des marchés pour se financer par crédit bancaire** plus sûr et devenu plus intéressant.

Des PPP sont également envisageables, par exemple la création d'une structure au capital mixte, une **"Société de Financement de la Transition Énergétique"** (SFTE)[15], sur le modèle de la Société de Financement de l'Économie Française (SFEF)[16]. De droit privé, la SFTE serait composée de différentes banques de crédit et pourrait lever des fonds sur les marchés financiers et auprès des investisseurs institutionnels pour financer des investissements verts massifs sans augmenter le service de la dette publique, mais bénéficierait d'une garantie publique, offrant des taux et un coût de financement bien plus faibles qu'aujourd'hui.

Une **politique de soutien à la participation des investisseurs institutionnels** au financement de la transition, qui s'incarnerait par la création de structures d'appariement public/privé hébergées par les banques publiques, apportant la connaissance et la centralisation de l'information nécessaires à la réunion projets/investisseurs[17]. Des structures privé/privé entre banques et investisseurs institutionnels pourraient aussi faciliter l'appariement.

Malgré une politique monétaire européenne, l'Etat peut être proactif à travers ses banques publiques, par exemple la Banque Publique d'Investissement et la Caisse des Dépôts et Consignations, en leur imposant l'arrêt du financement d'activités fossiles et une part plancher de financement des activités vertes, par exemple 60% (aujourd'hui aux alentours d'à peine 9%) et même une part fixe de la dette publique émise. **L'investissement public préliminaire et sa constance sont vitaux du fait du manque d'infrastructures initiales et sa versatilité.** En finançant de grands projets de transition, le public ouvrirait la voie en assumant les pertes initiales et en fournissant le capital-risque. De plus, en émettant des titres adossés à des obligations vertes, il habituerait et rassurerait les marchés à ces instruments.

---

[14] Fonds de pension, compagnies d'assurances, hedge funds et autres, avec un volume d'actifs gérés de 74 300 milliards USD (2018) qui permettrait de financer l'entière transition.
[15] G. Giraud, Collectif. (2015). *Rapport du Groupe de Travail n°4 du Débat National sur la Transition Énergétique.* DNTE.
[16] La SFEF a été créée en 2008 afin de sauver les banques françaises en levant 77 milliards d'euros.
[17] Voir par exemple la UK Pension Infrastructure Platform.

Enfin, l'échéance longue des investissements verts désincite les banques à les financer. L'Etat et la BPI pourraient créer un mécanisme de recyclage des emprunts à long terme en s'inspirant du mécanisme de refinancement de l'activité d'exportation des entreprises (cf. la COFACE), pour créer **une garantie publique de refinancement des emprunts à long terme** : l'Etat garantirait les fonds investis à hauteur de 100% à la fois en cas de défaut du projet mais aussi en cas de défaut de la banque privée prêteuse, si elle n'est pas en capacité de renouveler le crédit.

**Défossilisation de l'épargne**

Un dernier enjeu est la **"défossilisation" de l'épargne**. La quantité d'épargne existante est extrêmement importante (4000 milliards en France), et doit être massivement réallouée vers le financement de la transition. L'État pourrait imposer le **fléchage** de l'épargne vers la finance durable :

- Il pourrait être rendu obligatoire pour les banques et institutions financières de renseigner les épargnants sur l'utilisation de leur épargne et leur permettre des choix éthiques garantis par l'Etat. Cela nécessiterait d'imposer la généralisation de la notation socio-environnementale à tous les produits d'épargne.

- La France bénéficie déjà du Livret de développement durable et solidaire (LDDS), en 2021 à taux équivalent avec le Livret A (0,5%). Il est nécessaire de majorer le taux du LDDS pour créer une réelle incitation, et imposer légalement que l'épargne placée en LDDS soit intégralement investie dans des fonds labellisés, ce qui n'est toujours pas le cas aujourd'hui.

- **La fiscalité doit aussi orienter l'épargne**, en créant des avantages fiscaux pour les épargnants plaçant leurs fonds dans des produits labélisés, tout en conservant un niveau de recettes fiscales constant par la suppression de niches, exonérations et abattements concernant les activités et produits polluants. **Une trentaine d'exonérations françaises concernent des secteurs néfastes.**[18] Il est possible de taxer les polluants proportionnellement à leurs émissions, les gains permettant de compenser l'impact de la hausse du prix de l'essence sur les classes précaires.

- Les contraintes, coûts divers et difficultés techniques du *reporting* et de la notation pèsent particulièrement sur les petites et moyennes entreprises ainsi que sur les projets territoriaux. Aussi, l'Etat doit soutenir leur financement non-seulement par une garantie publique en prenant en charge leur solvabilité financière en contrepartie de leur solvabilité socio-environnementale, mais aussi en les aidant à l'organisation et à la syndication[19] pour obtenir ces fonds.

**Entreprise, Comptabilité et Solvabilité socio-environnementale**

Afin de répondre à notre double objectif, à savoir **faciliter l'emprunt vertueux** tout en limitant **les emprunts vers des activités négatives** et pour permettre la **garantie publique**, un système de notation des entreprises est nécessaire. Nous proposons donc **l'intégration de la solvabilité socio-environnementale dans l'analyse comptable des entreprises.**

---

[18] Les dépenses fiscales défavorables à l'environnement sont de 7,5 milliards en 2017 face à 3,1 milliards pour celles favorables à l'environnement.
[19] Centralisation des projets pour l'obtention d'un financement collectif.

Notamment, le **modèle CARE-TDL[20]** développé par J. Richard, propose l'intégration des capitaux humain et naturel aux côtés du capital financier au passif du bilan comptable. L'objectif d'une entreprise demeure la non-dégradation de son capital financier, mais la même logique s'applique également aux capitaux humain et naturel. L'entreprise dispose d'un écosystème qu'elle exploite pour créer de la valeur. Cependant, il ne doit pas être entièrement dégradé, quelle que soit l'utilisation que l'entreprise en fait.

**La méthode CARE a donc pour objectif de traduire une dette écologique que l'entreprise doit piloter afin d'assurer sa performance globale, maximisant ainsi sa solvabilité socio-environnementale et sa capacité de financement.** S'appuyant sur le principe de non-compensation comptable, elle valorise une approche en soutenabilité forte du développement durable. Soit, l'idée selon laquelle aucun capital n'est substituable à un autre, considérant la finitude des ressources et l'irréversibilité de la destruction de certains de leurs composants[21]. Par ailleurs, le modèle réaffirme l'importance du principe de sincérité de l'entreprise en institutionnalisant l'aléa moral dans la norme comptable, afin de lutter contre les pratiques de communication verte (*greenwashing*).

Toutefois, la notation socio-environnementale est un concept polysémique, tant au niveau des instruments de mesure et des dispositifs d'évaluation qu'au niveau des comportements des acteurs dont les décisions sont évaluées. La performance n'existe pas en elle-même : les systèmes d'évaluation sont basés sur les systèmes de valeurs de leurs concepteurs qui font appel à des principes et à des logiques de justification[22]. Ainsi, la question de la faisabilité de leur intégration aux normes comptables actuelles est techniquement posée et pour l'instant non résolue. Face à la limite de notre proposition, nous soulignons **la nécessité de créer une norme systémique internationale en termes de mesure d'impact socio-environnemental et ainsi pouvoir travailler sur des fondations communes vers l'Agenda 2030**.

**Conclusion**

Toute politique est un arbitrage coût-bénéfice. C'est particulièrement vrai pour les politiques prudentielles, car touchant à de nombreux intérêts et aux normes supranationales, et nécessitant d'être continuellement contrôlées et évaluées. De plus, la résistance au changement dans les activités hautement carbonées générant de hauts revenus est conséquente. **La puissance publique doit s'engager et envoyer des signaux forts**. La politique proposée dans cette note invite à la mise en place de mesures complémentaires, telles :

- La réglementation des *dark pools,* de leurs opérateurs et du trading à haute fréquence, voire le retrait d'accréditations. L'instauration réglementaire de chambres de compensation dans les marchés de gré à gré et d'un seuil-plancher pour les appels de marge des chambres afin de garantir leur coussin prudentiel, voire les renationaliser.

- L'investissement public vert massif, cohérent et continu, comme signal au secteur privé mais aussi pour activer des externalités positives par le développement des infrastructures.

L'application stricte du principe de pollueur-payeur, les amendes devant servir à la réparation socio-environnementale. Le développement de normes fortes, en menant une réelle pression pour une harmonisation fiscale et environnementale européenne afin d'éviter le *dumping* écologique (ie. la destruction de normes pour attirer des firmes transnationales).

---

[20] Richard, J. (2012). "Présentation du modèle CARE-TDL (Comptabilité Adaptée au Renouvellement de l'Environnement - Triple Depreciation Line)". *Comptabilité et Développement Durable*, ed. Economica.

[21] Daly, H.E., Cobb Jr, J. & Cobb, J. (1994). *For the common good: Redirecting the economy toward community, the environment, and a sustainable future,* Beacon Press.

[22] Brignall S., & Modell, S. (2000). "An institutional perspective on performance measurement and management in the new public sector". *Management Accounting Research*, Vol.11, p.281-306.